\documentclass[a4paper,11pt]{article}

\usepackage{pos}
\usepackage{lineno}
\usepackage{lipsum} 
\title{Observation of high-energy neutrinos from the \\ Galactic plane}

\ShortTitle{Observation of high-energy neutrinos from the Galactic plane}

\author{The IceCube Collaboration \\{\normalsize \normalfont(a complete list of authors can be found at the end of the proceedings)}\\}

\emailAdd{ssclafani@icecube.wisc.edu}
\emailAdd{mhuennefeld@icecube.wisc.edu}

\abstract{

IceCube has discovered a flux of astrophysical neutrinos and presented evidence for the first neutrino sources, a flaring blazar known as TXS 0506+056 and the active galaxy NGC 1068. However, the sources responsible for the majority of the astrophysical neutrino flux remain elusive. In addition to hypothetical sources within our Galaxy, high energy neutrinos are produced when cosmic rays interact at their acceleration sites and during propagation through the interstellar medium. The Galactic plane has therefore long been hypothesized as a neutrino source. In this contribution, new results are presented for searches of neutrino sources utilizing a dataset that builds upon recent advances in deep-learning-based reconstruction methods for neutrino-induced cascades. This work presents the first observation of high-energy neutrinos from the Milky Way Galaxy, rejecting the background-only hypothesis at 4.5~$\sigma$. The neutrino signal is consistent with diffuse emission from the Galactic plane, potentially in combination with emission by a population of sources.

\vspace{4mm}
{\bfseries Corresponding authors:}
Steve Sclafani$^{1*}$, Mirco H{\"u}nnefeld$^{2}$ \\
{$^{1}$ \itshape University of Maryland}\\
{$^{2}$ \itshape TU Dortmund University}\\[4mm]
$^*$ Presenter

\ConferenceLogo{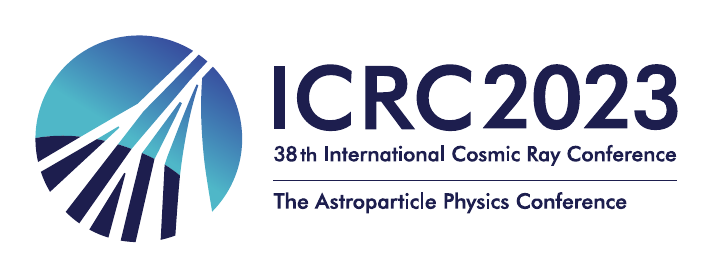}

\FullConference{The 38th International Cosmic Ray Conference (ICRC2023)\\ 26 July -- 3 August, 2023\\ Nagoya, Japan}
}

\begin{document}

\maketitle

\section{Neutrino Production in the Milky Way}\label{intro}

The Milky Way Galaxy has been observed over many wavelengths of light, from radio waves to gamma rays. At energies above 1\,GeV, the sky as observed by \textit{Fermi}-LAT \cite{fermi} is dominated by emission from the Galactic plane of the Milky Way.  Cosmic rays can interact at their acceleration site creating high-energy neutrinos and gamma rays, but can also escape the acceleration environment.  The cosmic rays will generally diffuse through the Galaxy, and may interact with the dust and gas of the plane, producing emission at the site of these interactions.  The observed emission reflects the density and energy of cosmic rays, as well as the column density of matter.  This diffuse emission has been observed in gamma rays \cite{fermi, lhaaso, tibet}, and the counterpart has been predicted in high-energy neutrinos \cite{Stecker1979, BEREZINSKY1993}.   This work presents the first evidence of high-energy neutrinos from the Milky Way Galaxy \cite{sciencearticle}.  

The IceCube Neutrino Observatory, a cubic-kilometer detector located in the geographic South Pole, is instrumented with 5160 Digital Optical Modules (DOMs), each with a PMT and digitizing hardware.  When a neutrino interacts with the ice, charged secondary particles produce Cherenkov light in the detector that can be used to reconstruct the direction and energy of the primary neutrino.  These interactions occur in two main topologies.  ``Track'' events are mostly produced from ``charged-current'' (CC) $\nu_\mu$ interactions that produce an outgoing muon, which traverses the detector and deposits light along the particle track, whereas ``cascade'' events are produced from ``neutral-current'' (NC) interactions of all flavors, as well as CC$-\nu_e$ or CC-$\nu_\tau$ interactions.  These produce a shower of particles that appear as a nearly spherically expanding light front.  For this reason, typical cascade events have inferior angular resolution (\textgreater 10$^\circ$ at 10 TeV) when compared to track events (\textless 1$^\circ$).  After IceCube discovered a flux of astrophysical neutrinos \cite{astro}, it used track events to present evidence for two sources \cite{txs_precurser, ngc}.  The sources responsible for the majority of the astrophysical flux, however, remain unknown.  The diffuse flux from the Galactic plane can account for up to $\sim$10\% of this flux at TeV energies~\cite{icecube_antares}.

Previous searches for emission from the Galactic plane did not find any significant emission~\cite{gptracks,icecube_antares, mesecascades, antares}.  The level of Galactic emission is expected to be low when compared to the measured all-sky astrophysical flux, and due to the direction of the Galactic center, diffuse emission is concentrated in the southern sky.  For IceCube, due to the location of the detector at the South Pole,  selections in the southern celestial sky are composed of events downgoing in the detector.  Searches in this region are particularly difficult due to the large background of atmospheric muons. Through-going track-based analyses see a reduction of sensitivity due to this irreducible background~\cite{tracks10yr}.  This is especially true for galactic sources, which are assumed to follow a softer spectrum.  This work, however, makes use of cascade events, which have a much reduced background in the southern sky, and lead to an improvement of sensitivity.  Further, machine learning techniques applied to IceCube cascade events improve the sensitivity by a factor of 3-4 over previous searches~\cite{sciencearticle}.  Consequently, this work improves sensitivity to diffuse Galactic plane emission models, leading to the first evidence of high-energy neutrino emission from the Milky Way at a significance of 4.5$\sigma$.  

\section{Application of Deep Learning to Cascade Events}

\begin{figure}
    \centering
    \includegraphics[width=\textwidth]{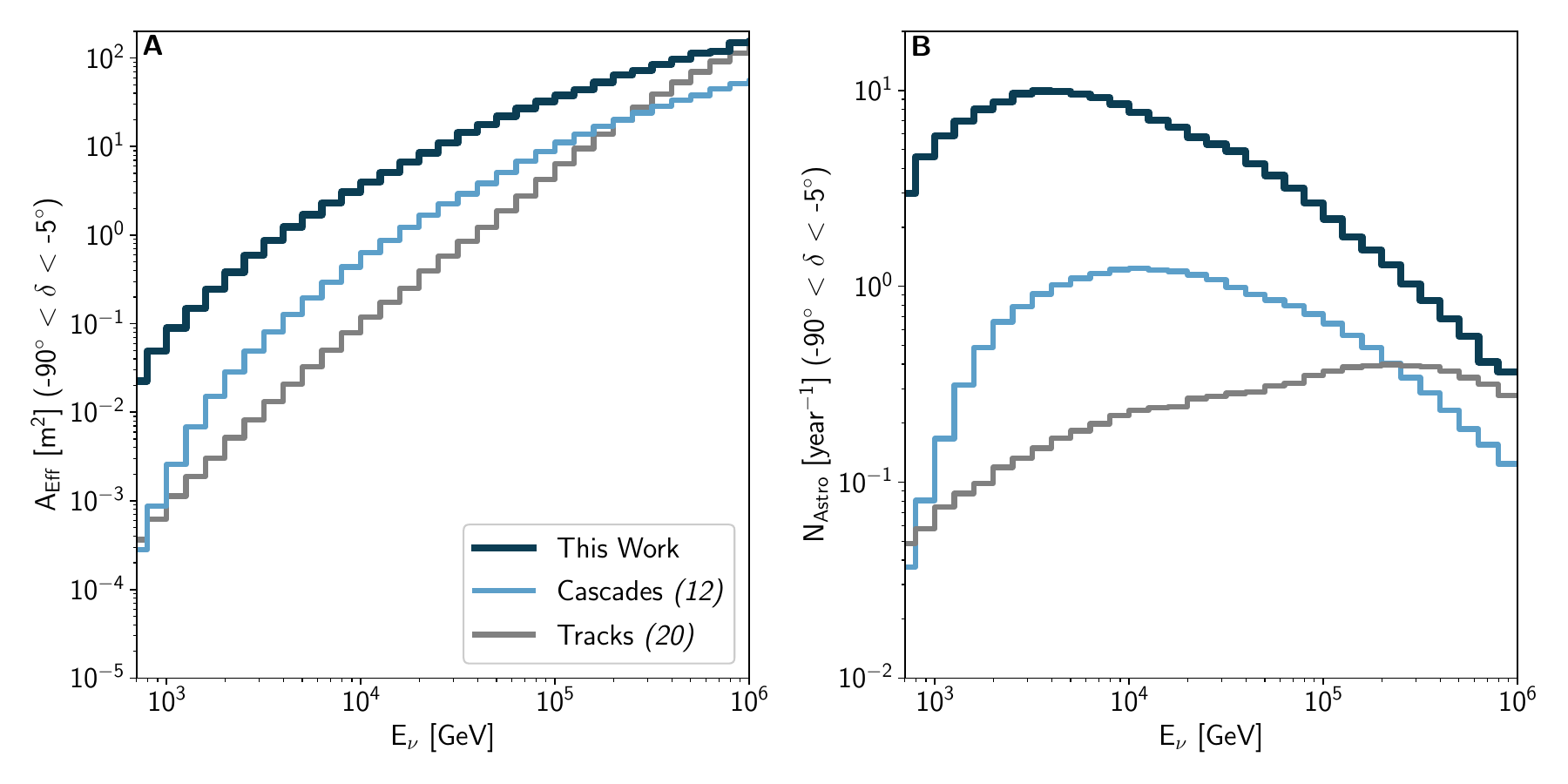}
    \caption{ (A) The all-flavor southern sky effective area (A$_{\rm Eff}$) of the IceCube dataset, averaged over solid angle in the declination ($\delta$) range between $-90^\circ$ and $-5^\circ$ as a function of E$_\nu$, the true neutrino energy.  Results are shown for the deep learning event selection used in this work, (dark blue), a previous cascade event selection~\cite{mesecascades} (light blue), and , as an example, a previous though-going track event selection~\cite{tracks10yr} (grey) applied to the IceCube data. (B) The number of expected signal events (N$_{Astro}$) in the southern sky per energy bin per year for each event selection, assuming an isotropic astrophysical flux~\cite{diffuse_cascades}. Calculations are based on equal contributions of each neutrino flavor at Earth due to neutrino oscillations, (figure from~\cite{sciencearticle})}
    \label{fig:effa}
\end{figure}

In the southern sky, the primary background for IceCube consists of atmospheric muons, which trigger the detector at a rate a billion times higher than astrophysical neutrinos. This background appears in the detector as downgoing track events which pass through the detector.  These events cannot be distinguished from neutrino-induced muons entering the detector. Thus, the event selection in this work is optimized to search for events that start within the detector and have cascade-like topologies.  The selection of cascade events instead of track events reduces the atmospheric muon background, and also suppresses the atmospheric neutrino background, which primarily consists of muon flavored neutrinos.   Previous cascade searches \cite{mesecascades} used an event selection based on veto layers that improved the effective area over tracks in the 1-100 TeV range~\cite{mese}.  This work, however, relies heavily on a Deep-Neural-Network (DNN) based selection~\cite{CNNPaper} that improves the selection efficiency \cite{sciencearticle} at low energies.  Further, this work includes events that are partially contained within the detector, which improves the efficiency mostly at higher energies.  Due to the reduced backgrounds, the lower energy threshold is pushed lower, from a tens of TeV down to around 1 TeV. The effective area at 10 TeV is increased by a factor of $\sim$5 when compared to previous cascade selections, as shown in Figure \ref{fig:effa}A.  This increase in effective area corresponds to a an increase of astrophysical events by up to a factor of 20, assuming the measurement in \cite{diffuse_cascades}, as shown in Figure \ref{fig:effa}B. Relative improvements are largest at energies below a few tens of TeV.

The event selection also relies on a novel reconstruction that combines the advantages of DNNs with traditional maximum likelihood methods~\cite{event-generator}. The new reconstruction technique results in a similar, sample-averaged angular resolution at high energies as obtained by the previous cascade sample, despite the inclusion of more challenging events. At low energies, the sample-averaged resolution is improved by up to a factor of two. These improvements are achieved due to better exploitation of available information and symmetries in the novel reconstruction method~\cite{event-generator}.

\section{Searches for Galactic neutrino emission}

Multiple source hypotheses were tested.  The focus of this work is the testing of three models of diffuse Galactic plane emission.  The first model, referred to here as $\pi^0$, is based on the extrapolation from the \textit{Fermi}-LAT fit to the first 21 months of data \cite{fermi}.  The spatial template from \textit{Fermi}-LAT is assumed unchanged at TeV energies, and the energy spectrum is assumed to be an unbroken power-law with fixed spectral index $\gamma$=2.7.
The other two models, referred to as KRA$_\gamma$, model the Galactic cosmic ray diffusion coefficient as radially dependent \cite{kra}. This results in a harder spectrum towards the galactic center which then results in an enhanced flux after extrapolation to IceCube's energies. Here we test the extrapolated flux, but use a global energy dependant spectrum, averaged over all directions. Two cutoffs are tested. KRA$_\gamma^{5}$ and KRA$_\gamma^{50}$ correspond to cutoffs in Galactic cosmic rays at 5 PeV and 50 PeV respectively.  These energies are propagated down to lower energies in neutrinos.  The sky-integrated flux of these models are shown in Figure \ref{fig:fluxnorm}.

For each template, the number of signal events is fitted while the spectrum is held fixed. A likelihood-ratio test is performed with the log-likelihood ratio as the test statistic (TS)~\cite{gptracks}.  Background TS distributions are calculated from scrambled data, and the true TS for each template is compared with this background distribution to calculate the p-value.  Due to IceCube's location on the Earth, scrambling the Right Ascension (RA) direction produces background-like pseudo-experiments, but preserves any detector effects.  Using this data driven technique to derive p-values, any unmodeled effect will result in a reduction of sensitivity, but not risk a false signal.

In addition to the diffuse Galactic plane analyses, searches were performed for neutrino emission from Galactic source classes corresponding to supernova remnants (SNR), pulsar wind nebulae (PWN) and unidentified TeV gamma-ray sources (UNID).  Stacking sources can result in an improved sensitivity over searching for each source individually. Three classes of 12 potential Galactic sources are selected based on their TeV gamma-ray flux, and each is source is weighted to equally contribute.  The complete source list is available in Ref. \cite{sciencearticle}.

\section{Neutrinos from the Milky Way}\label{results}

The results for the diffuse Galactic plane and source stacking are shown in Table \ref{tab:results} and Table \ref{tab:stacking} respectively, which was first reported in \cite{sciencearticle}. Each of the diffuse emission hypotheses is significant with pre-trial p-values corresponding to 4.71$\sigma$, 4.37$\sigma$ and 3.96$\sigma$, respectively.  These fluxes correspond to  best-fitting values of 748, 276, and 211 signal events ($n_s$) in the IceCube dataset for the $\pi^0$, KRA$_\gamma^5$ and KRA$_\gamma^{50}$ models, respectively. After accounting for the three tested hypotheses, the resulting post-trials p-value is calculated,and coresponds to a significance of 4.5$\sigma$.

\begin{table}[ht]
    \centering
    \begin{tabular}{ccc cc}
    \hline
         $\begin{array}{cc}
              \textbf{Diffuse Galactic} \\
              \textbf{plane analyses} 
         \end{array}$ & 
         $\begin{array}{cc}
              \textbf{Flux Sensitvity} \\
              \textbf{$\Phi$} 
         \end{array}$ & 
         $\begin{array}{cc}
              \textbf{Best-fitting} \\
              \textbf{$n_s$} 
         \end{array}$ & 
         \textbf{p-value}   & 
         $\begin{array}{cc}
              \textbf{Best-fitting} \\
              \textbf{flux $\Phi$} 
              \end{array}$ \\
         \hline
    \rule{0pt}{2ex} $\pi^0$ &  5.98 &  748 &  1.3$\times$10$^{-6}$  (4.71$\sigma$) &  21.8 $^{+5.3}_{-4.9}$  \\
    \rule{0pt}{2ex}KRA$_\gamma^{5}$ & 0.16$\times$MF &  276 & 6.1$\times$10$^{-6}$  (4.37$\sigma$) &   0.55$^{+0.18}_{-0.15}\times$MF \\
    \rule{0pt}{2ex}KRA$_\gamma^{50}$ & 0.11$\times$MF & 211 & 3.7$\times$10$^{-5}$ (3.96$\sigma$) &  0.37$^{+0.13}_{-0.11}\times$MF \\
    \hline
    \end{tabular}
    \caption{ The flux sensitivity and best-fitting flux normalization ($\Phi$) are given in units of model flux (MF) for KRA$_\gamma$ templates and as E$^2$ $\frac{dN}{dE}$ at 100\,TeV in units of 10$^{-12}$ TeV\,cm$^{-2}$\,s$^{-1}$ for the $\pi^0$ analyses ($\frac{dN}{dE}$ is the differential number of neutrinos per flavor, N, and neutrino energy, E). P-values and significance are calculated with respect to the background-only hypothesis. Pre-trial p-values for each individual result are shown for the three diffuse Galactic plane analyses.}

    \label{tab:results}
\end{table}

\begin{figure}
    \centering
    \includegraphics[width=0.7\textwidth]{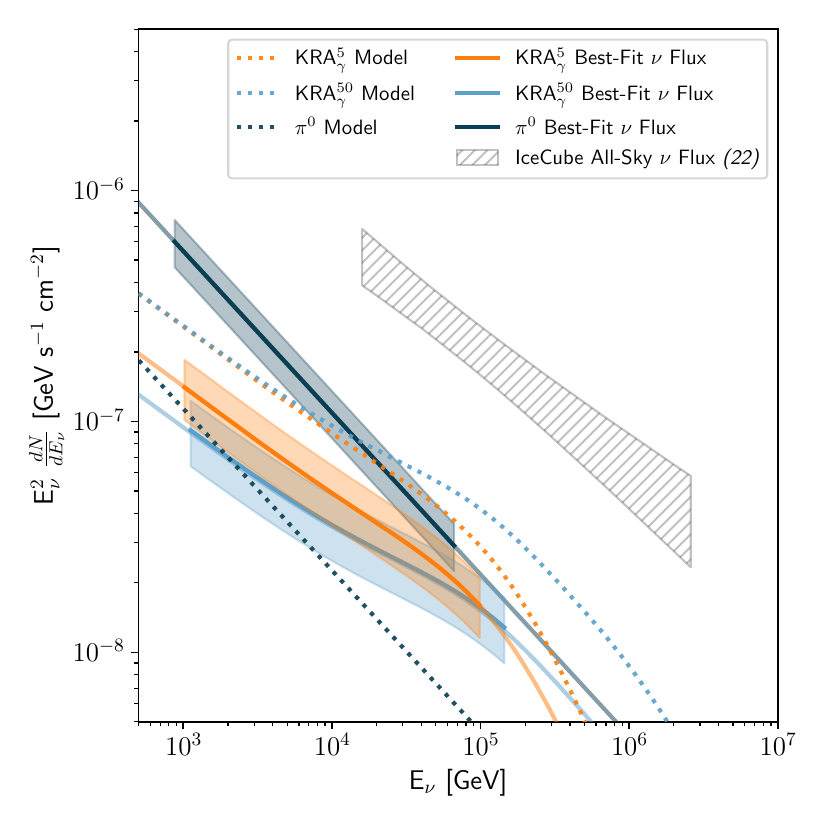}
    \caption{Energy-scaled, sky-integrated, per-flavor neutrino flux as a function of neutrino energy (E$_\nu$) for each of the Galactic plane models. Dotted lines are the predicted values for the $\pi^0$ (dark blue), KRA$_\gamma^5$ (orange) and KRA$_\gamma^{50}$ (light blue) models while solid lines are our best-fitting flux normalizations from the IceCube data. Shaded regions indicate the 1$\sigma$ uncertainties, extending over the energy range that contributes to 90$\%$ of the significance.  These results are based on the all-sky (4$\pi$ sr) template and are presented as an all-sky flux. For comparison, the grey hatching shows the flux of the IceCube all-sky neutrino flux~\cite{diffuse_cascades}, scaled to an all-sky flux by multiplying by 4$\pi$, with its 1$\sigma$ uncertainty. Figure from~\cite{sciencearticle}}
    \label{fig:fluxnorm}
\end{figure}

Also included in Table \ref{tab:results} are best-fit flux normalizations for each template.  The $\pi^0$ flux is reported at 100 TeV.  Since the KRA$_\gamma$ models do not follow a power-law, the flux is reported as multiples of the model prediction.   Those normalizations are shown compared to the model for each template in Figure \ref{fig:fluxnorm}.  The KRA$_\gamma$ best-fitting normalization is less than the models, whereas the $\pi^0$ best-fit is larger by a factor of $\sim$5 than the $\pi^0$ model. This could be an indication that the underlying diffuse emission spectrum is more complex than assumed in any of the tested models, possibly containing a contribution from unresolved Galactic sources.  These results are based on the all-sky template and model-to-model flux comparisons depend on the considered sky region.  When considering regions of the sky tested by gamma-ray experiments like Tibet-AS$\gamma$ the best-fit $\pi^0$ model fits well with the measured gamma-ray flux.  The IceCube result, however, is a measurement over the entire sky and it does not attempt to mask out any galactic sources, as none have been identified.

In addition to the Galactic diffuse templates, the three Galactic stacking analyses also resulted in evidence above 3$\sigma$ with pre-trial significance of 3.24$\sigma$, 3.24$\sigma$, and 3.40$\sigma$ for the SNR, PWN, and UNID classes respectively.  
The full results are shown in Table~\ref{tab:stacking}.  
Due to the large spatial overlap in source hypotheses, these significances are expected in the scenario of diffuse emission at the best-fit values reported in Table~\ref{tab:results}. 
This analysis cannot distinguish the contribution of individual sources or classes of sources from diffuse emission. 

\begin{table}
    \centering

    \vspace{0.5em}
    \begin{tabular}{cccccccc}
    \hline
         Catalog & Sensitivity $\Phi$ & n$_s$ &  $\gamma$ & p-value & Significance ($\sigma$) & Flux $\Phi$ & UL $\Phi$ \\
         \hline
         SNR & 2.24 & 218.6 &  2.75 & 5.9$\times$10$^{-4}$ & 3.24 & 6.22 & \textless9.01 \\
         PWN & 2.25 & 279.6 & 3.00 & 5.9$\times$10$^{-4}$ & 3.24 & 3.80 & \textless9.50 \\
         UNID & 1.89 & 238.4 & 2.85 & 3.4$\times$10$^{-4}$ & 3.40 & 5.03 & \textless7.76 \\
         \hline
    \end{tabular}
    \caption{ Pre-trial significance, sensitivity, best-fitting spectrum ($\gamma$), total number of signal events (n$_s$), flux, and 90\% Upper Limits (UL) for the Galactic stacking catalog analyses.  Upper limits are with respect to a source emitting following an E$^{-2}$ spectrum. 
    The per-flavor neutrino flux sensitivity, best-fit, and upper limits are given as E$^2$ $\frac{dN}{dE}$ at 100\,TeV in units of 10$^{-12}$ TeV\,cm$^{-2}$\,s$^{-1}$ for the entire catalog of sources.}
    \label{tab:stacking}
\end{table}

\begin{figure}
    \centering
    \includegraphics[width=0.9\textwidth]{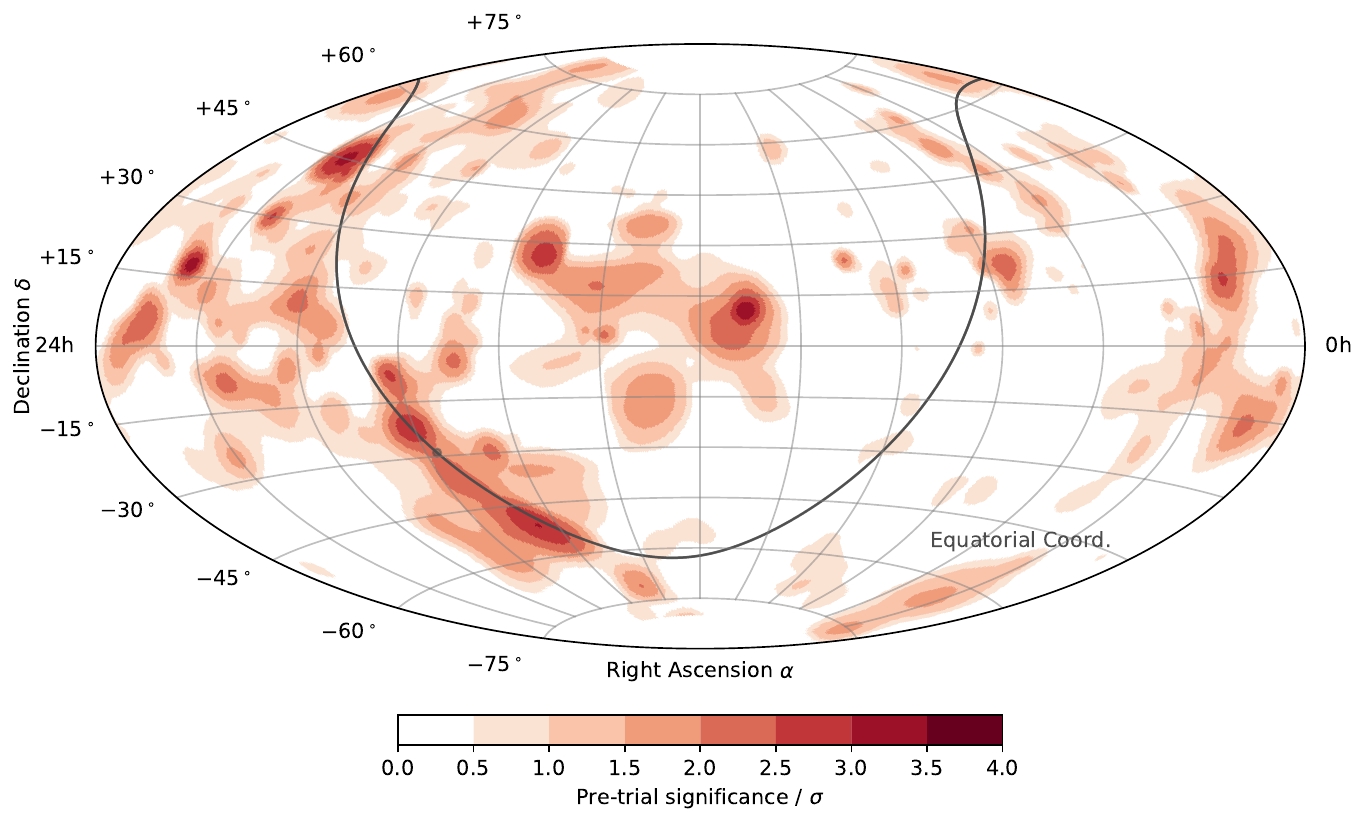}
    \caption{The best-fitting pre-trial significance for the all-sky search is shown as a function of direction in an Aitoff projection of the celestial sphere, in equatorial coordinates (J2000 equinox). The Galactic plane is indicated by a grey curve, and the Galactic Center as a dot. Although some locations appear to have significant emission, the trial factor for the number of points searched means these points are all individually statistically consistent with background fluctuations. The clustering of larger significances along the galactic plane reflects the significant excess that is observed in the template searches for the Galactic plane. Figure from~\cite{sciencearticle}. }
    \label{fig:skymap}
\end{figure}

Finally, an all-sky scan was performed, searching at each location on the sky for an excess of point-like neutrino emission. The visualization of this all-sky search is shown in equatorial coordinates in Figure \ref{fig:skymap}.  Excesses are visible along the Galactic plane and near well known gamma-ray emitters, such as the Crab Nebula, 3C 454.3, and the Cygnus X region, however, no single point is significant once trials are accounted for.

\section{Conclusion}\label{conclusion}

This work presents the first evidence of high-energy neutrino emission from the Milky Way Galaxy. The result is consistent with a diffuse neutrino emission hypothesis or a collection of unresolved sources.  

Round-trip tests were performed to investigate the correlation between the source stacking analyses and the diffuse Galactic plane analyses.  Injecting the best-fitted flux from the $\pi^0$ template results in a significance for the other templates that is compatible with the observed results.  Injecting any one of the stacking source hypotheses individually, does not recover the best-fitted flux for the diffuse emission models.  Further, it is possible that some of these sources still contribute to the observed flux.

More information is needed to characterize the Galactic component of the astrophysical neutrino flux.  A larger range of Galactic templates, including more recent models \cite{Schwefer_2023}, have been tested with a dataset of northern sky track events in IceCube \cite{ntgp_icrc}. A search for the same diffuse templates was also performed with IceCube starting track events \cite{estes_icrc,Aartsen:2023prd:estes}. Both results are compatible with the results presented in this work.  Further work will require the identification of sources and their spectra.  This could be done by combining the signal purity of cascade events with the angular resolution of track events \cite{combo_icrc}.

\bibliographystyle{ICRC}
\bibliography{references}

%

\clearpage

\section*{Full Author List: IceCube Collaboration}

\scriptsize
\noindent
R. Abbasi$^{17}$,
M. Ackermann$^{63}$,
J. Adams$^{18}$,
S. K. Agarwalla$^{40,\: 64}$,
J. A. Aguilar$^{12}$,
M. Ahlers$^{22}$,
J.M. Alameddine$^{23}$,
N. M. Amin$^{44}$,
K. Andeen$^{42}$,
G. Anton$^{26}$
% ...
\\
\\
% ...
$^{12}$ Universit{\'e} Libre de Bruxelles, Science Faculty CP230, B-1050 Brussels, Belgium \\
% ...
$^{17}$ Department of Physics, Loyola University Chicago, Chicago, IL 60660, USA \\
$^{18}$ Dept. of Physics and Astronomy, University of Canterbury, Private Bag 4800, Christchurch, New Zealand \\
% ...
$^{22}$ Niels Bohr Institute, University of Copenhagen, DK-2100 Copenhagen, Denmark \\
$^{23}$ Dept. of Physics, TU Dortmund University, D-44221 Dortmund, Germany \\
% ...
$^{26}$ Erlangen Centre for Astroparticle Physics, Friedrich-Alexander-Universit{\"a}t Erlangen-N{\"u}rnberg, D-91058 Erlangen, Germany \\
% ...
$^{40}$ Dept. of Physics and Wisconsin IceCube Particle Astrophysics Center, University of Wisconsin{\textendash}Madison, Madison, WI 53706, USA \\
% ...
$^{42}$ Department of Physics, Marquette University, Milwaukee, WI, 53201, USA \\
% ...
$^{44}$ Bartol Research Institute and Dept. of Physics and Astronomy, University of Delaware, Newark, DE 19716, USA \\
% ...
$^{63}$ Deutsches Elektronen-Synchrotron DESY, Platanenallee 6, 15738 Zeuthen, Germany  \\
$^{64}$ Institute of Physics, Sachivalaya Marg, Sainik School Post, Bhubaneswar 751005, India \\
% ...

\subsection*{Acknowledgements}

\noindent
The authors gratefully acknowledge the support from the following agencies and institutions:
USA {\textendash} U.S. National Science Foundation-Office of Polar Programs,
U.S. National Science Foundation-Physics Division,
U.S. National Science Foundation-EPSCoR,
Wisconsin Alumni Research Foundation,
Center for High Throughput Computing (CHTC) at the University of Wisconsin{\textendash}Madison,
Open Science Grid (OSG),
Advanced Cyberinfrastructure Coordination Ecosystem: Services {\&} Support (ACCESS),
Frontera computing project at the Texas Advanced Computing Center,
U.S. Department of Energy-National Energy Research Scientific Computing Center,
Particle astrophysics research computing center at the University of Maryland,
Institute for Cyber-Enabled Research at Michigan State University,
and Astroparticle physics computational facility at Marquette University;
Belgium {\textendash} Funds for Scientific Research (FRS-FNRS and FWO),
FWO Odysseus and Big Science programmes,
and Belgian Federal Science Policy Office (Belspo);
Germany {\textendash} Bundesministerium f{\"u}r Bildung und Forschung (BMBF),
Deutsche Forschungsgemeinschaft (DFG),
Helmholtz Alliance for Astroparticle Physics (HAP),
Initiative and Networking Fund of the Helmholtz Association,
Deutsches Elektronen Synchrotron (DESY),
and High Performance Computing cluster of the RWTH Aachen;
Sweden {\textendash} Swedish Research Council,
Swedish Polar Research Secretariat,
Swedish National Infrastructure for Computing (SNIC),
and Knut and Alice Wallenberg Foundation;
European Union {\textendash} EGI Advanced Computing for research;
Australia {\textendash} Australian Research Council;
Canada {\textendash} Natural Sciences and Engineering Research Council of Canada,
Calcul Qu{\'e}bec, Compute Ontario, Canada Foundation for Innovation, WestGrid, and Compute Canada;
Denmark {\textendash} Villum Fonden, Carlsberg Foundation, and European Commission;
New Zealand {\textendash} Marsden Fund;
Japan {\textendash} Japan Society for Promotion of Science (JSPS)
and Institute for Global Prominent Research (IGPR) of Chiba University;
Korea {\textendash} National Research Foundation of Korea (NRF);
Switzerland {\textendash} Swiss National Science Foundation (SNSF);
United Kingdom {\textendash} Department of Physics, University of Oxford.

\end{document}